# Dimension and Relative Frequencies

Xiaoyang Gu*
Department of Computer Science
Iowa State University
Ames, IA 50011 USA
xiaoyang@cs.iastate.edu

Jack H. Lutz*†
Department of Computer Science
Iowa State University
Ames, IA 50011 USA
lutz@cs.iastate.edu

**Abstract**

We show how to calculate the finite-state dimension (equivalently, the finite-state compressibility) of a saturated sets $X$ consisting of *all* infinite sequences $S$ over a finite alphabet $\Sigma_m$ satisfying some given condition $P$ on the asymptotic frequencies with which various symbols from $\Sigma_m$ appear in $S$. When the condition $P$ completely specifies an empirical probability distribution $\pi$ over $\Sigma_m$, i.e., a limiting frequency of occurrence for *every* symbol in $\Sigma_m$, it has been known since 1949 that the Hausdorff dimension of $X$ is precisely $\mathcal{H}(\pi)$, the Shannon entropy of $\pi$, and the finite-state dimension was proven to have this same value in 2001.

The saturated sets were studied by Volkmann and Cajar decades ago. It got attention again only with the recent developments in multifractal analysis by Barreira, Saussol, Schmeling, and separately Olsen. However, the powerful methods they used – ergodic theory and multifractal analysis – do not yield a value for the finite-state (or even computable) dimension in an obvious manner.

We give a pointwise characterization of finite-state dimensions of saturated sets. Simultaneously, we also show that their finite-state dimension and strong dimension coincide with their Hausdorff and packing dimension respectively, though the techniques we use are completely elementary. Our results automatically extend to less restrictive effective settings (e.g., constructive, computable, and polynomial-time dimensions).

## 1 Introduction

Since the study of normal numbers by Borel in 1909 [4], sets of real numbers specified in terms involving frequencies of digits in their expansions have been repeatedly under the spotlight. Borel's results show that the set of real numbers for which the asymptotic frequencies of digits in the expansion follow the uniform distribution (normal numbers) has measure 1. However, this means that all sets of real numbers defined with any other asymptotic frequencies are measure 0 sets. Hence Lebesgue measure could not give more insight into such sets.

The introduction of Hausdorff dimension in 1919 [15] provides us a powerful tool to analyze the relative size of measure 0 sets and their geometry. This also enabled the study of sets of non-normal numbers defined by asymptotic frequencies of digits. Besicovitch (1934) [3] proved that in Cantor space **C**

$$\dim_{\mathrm{H}}(\mathrm{FREQ}^{\leq\beta}) = \mathcal{H}_2((\beta, 1-\beta)), \tag{1.1}$$

where $\beta \in [0, \frac{1}{2}]$, $\mathrm{FREQ}^{\leq\beta} = \{S \in \mathbf{C} \mid \limsup_{n \to \infty} \pi_0(S, n) \leq \beta\}$, $\pi_0(S, n)$ is the frequency of the occurrences of 0 in the first $n$ bits of $S$, and $\mathcal{H}_2(\cdot)$ is the binary entropy function. Then in 1941, Good [12] conjectured that

$$\dim_{\mathrm{H}}(\mathrm{FREQ}_\beta) = \mathcal{H}_2((\beta, 1-\beta)),$$

---
*This research was supported in part by National Science Foundation Grants 9988483 and 0344187. Part of the results were announced (without proceedings) at the special session on Randomness in Computation in Fall 2005 Central Section Meeting of the American Mathematical Society.
†This research is supported in part by Spanish Government MEC Projects TIC 2002-04019-C03-03 and TIN 2005-08832-C03-02.

which was only proven in 1949 by Eggleston [9], where $\text{FREQ}_\beta = \{S \in \mathbf{C} \mid \lim_{n\to\infty} \pi_0(S, n) = \beta\}$. It is worth noting that these are also among the earliest attempts to study the dimension structures of non-compact sets.

In 2000, Lutz gave a gale characterization of Hausdorff dimension and generalized Hausdorff dimension to resource-bounded and constructive dimensions [17, 18] and finite-state dimension [8] (with Dai, Lathrop, Mayordomo). For any set $X$, the dimensions following the inequality

$$\dim_{\text{FS}}(X) \geq \dim_\Delta(X) \geq \text{cdim}(X) \geq \dim_{\text{H}}(X),$$

where $\Delta$ is some computable resource bound. In [8], the finite-state dimensions of the above sets are shown to be the same as their Hausdorff dimensions.

Volkmann [30] and his student Cajar in his 1981 Ph.D. thesis [5] investigated more sets of this kind, with the central component of the definition of the sets being that they are *saturated*, as they are the sets that contain *all* the real numbers with some prescribed restrictions on the asymptotic behavior of the frequencies of digits. Cajar called these sets *saturated sets*. Volkmann [30] established that the Hausdorff dimension of saturated sets follow the maximum entropy principle, i.e., the Hausdorff dimension takes the value of the maximum entropy of the frequency allowed by the constraints in the definition of the saturated set.

It is important to note that the set $\text{FREQ}^{\leq\beta}$ exhibits the decomposition

$$\text{FREQ}^{\leq\beta} = \bigcup_{\beta'\leq\beta} \{S \in \mathbf{C} \mid \limsup_{n\to\infty} \pi_0(S, n) = \beta'\}.$$

Denote $\{S \in \mathbf{C} \mid \limsup_{n\to\infty} \pi_0(S, n) = \beta'\}$ as $\Delta_{\beta'}$. Each component $\Delta_{\beta'}$ has Hausdorff dimension

$$\dim_{\text{H}}(\Delta_{\beta'}) = \mathcal{H}_2((\beta', 1-\beta')).$$

Now we have that

$$\dim_{\text{H}}(\text{FREQ}^{\leq\beta}) = \sup_{\beta'\leq\beta} \dim_{\text{H}}(\Delta_{\beta'}).$$

The dimension magically exhibits stability under the *uncountable* union of the decomposition, which is not a typical behavior of the Hausdorff dimension, since it is only countably stable in general. The decomposition here is not artificially chosen to make the uncountable stability happen. This is the standard decomposition done in multifractal analysis to study the properties of the level sets in the fractal spectrum. The phenomenon of this uncountable stability was observed by Cajar from earlier works along this line [3, 16, 9, 10, 29, 30, 6] and led to the his own investigation [5]. However, besides Volkmann and Cajar, it was not until the 21st century, after the development of multifractal analysis, that this phenomenon was investigated again. Barreira, Saussol, and Schmeling, in their investigation of higher-dimensional multifractal analysis of some hyperbolic dynamical systems, obtained a conditional variational principle that allows the use of the entropy of ergodic measures to approximate topological entropy of the level set in the multifractal spectrum. Subsequently, they applied that to calculate the Hausdorff dimensions of many saturated sets. Olsen and Winter used techniques from large deviation theory to develop a unifying multifractal framework based on the notion of deformations of empirical measures [25, 26, 20, 24]. Olsen used this new framework to study the saturated sets [22, 23] and obtained the Hausdorff dimensions of a wide variety of such sets and some packing dimension results. And he mentioned the new multifractal framework as the main source of methods and techniques he used to solve the uncountable union problem [23].

Olsen noted in [21] that the sets of points whose limit of asymptotic frequencies does not exist received little attention in the past except for Volkmann and Cajar until the very recent work on the divergence points in multifractal analysis. In this paper, we will further this line of research by extending the results about saturated sets, especially, such sets of points whose asymptotic frequencies do not exist, to finite-state dimensions. We establish a pointwise characterization of the finite-state dimensions of these sets in terms of the asymptotic entropies, namely,

$$\dim_{\text{FS}}(X) = \sup_{S\in X} \liminf_{n\to\infty} \mathcal{H}(\vec{\pi}(S, n))$$

for all saturated sets $X$, where $\vec{\pi}(S,n) \in \Delta(\Sigma_m)$ is the empirical probability measure drawn from the frequencies of the occurrences of digits in the first $n$ bits of $S$. At the same time, we also establish a correspondence principle for Hausdorff dimension of saturated sets, namely,

$$\dim_{\text{FS}}(X) = \dim_{\text{H}}(X).$$

Packing dimension is another tool in classical geometric measure theory [28, 27]. In 2004, Athreya, Hitchcock, Lutz, and Mayordomo gave a very *surprising* gale characterization of packing dimension and generalized the packing dimension to resource-bounded, constructive, and finite-state strong dimension [1]. In this paper, we also investigate the packing dimension and finite-state strong dimension. We establish the finite-state strong dimension and packing dimension versions of exact duals of our aforementioned results in finite-state dimension and Hausdorff dimension, namely,

$$\dim_{\text{P}}(X) = \text{Dim}_{\text{FS}}(X) = \sup_{S \in X} \limsup_{n \to \infty} \mathcal{H}(\vec{\pi}(S,n)),$$

for any saturated set $X$. Among all the previously mentioned works, packing dimensions results are fewer. This is a strong indication that the gale characterization is particularly useful in the sense that for all the results concerning Hausdorff dimension and finite-state dimension in this paper, by using the gale characterization, we immediately obtain dual results about packing dimension and finite-state strong dimension using symmetric arguments. Also we use only elementary techniques to prove results that classically needed advanced techniques in ergodic theory and multifractal analysis. This fact suggests that gale characterization of dimensions and the finite-state dimension are very powerful.

Section 2 lists the basic definitions and conventions we use in this paper. Section 3 reviews the definitions of Hausdorff dimension, packing dimension, finite-state dimension, and finite-state strong dimension. We give a few example of calculating the dimensions of exotic saturated sets in Section 4. In Section 5, we discuss finite-state dimensions of saturated sets in detail and give insight into why a maximum entropy principle holds.

## 2 Preliminaries

Let $m \geq 2$ be an integer. We work with the $m$-ary alphabet $\Sigma_m = \{0, 1, \ldots, m-1\}$. $\Sigma_m^*$ is the set of all (finite) *strings* on $\Sigma_m$ including the empty string $\lambda$. $\mathbf{C}_m = \Sigma_m^\infty$ is the of all (infinite) $m$-ary *sequences*. $\mathbf{C} = \mathbf{C}_2$ is the Cantor space. $\Delta(\Sigma_m)$ is the set of all probability measures on $\Sigma_m$.

Let $i$ be an integer such that $0 \leq i \leq m-1$. The symbol counting function $\#_i : (\mathbf{C}_m \cup \Sigma_m^*) \times \mathbb{N} \to \mathbb{N}$ is defined such that for every string or sequence $S$ and $n \in \mathbb{N}$, $\#_i(S,n)$ is the number of occurrences of $i$ in the first $n$ bits of $S$. The symbol frequency function $\pi_i : (\mathbf{C}_m \cup \Sigma_m^*) \times \mathbb{N} \to [0,1]$ is defined such that $\pi_i(S,n) = \#_i(S,n)/n$. The empirical measure function $\vec{\pi} : (\mathbf{C}_m \cup \Sigma_m^*) \times \mathbb{N} \to \Delta(\Sigma_m)$ is defined such that $\vec{\pi}(S,n) = (\pi_0(S,n), \ldots, \pi_{m-1}(S,n))$. Intuitively, $\vec{\pi}$ extracts empirical probability measures from the first $n$ bits of a string or a sequence based on the actual frequencies of digits.

## 3 The Four Dimensions

Hausdorff dimension and packing dimension are important tools in mathematics used to study the size of sets and the properties of dynamic systems. All countable sets have 0 for both of these dimensions. In order to study relative size of countable sets from the eyes of computers with different resources, Lutz generalized Hausdorff dimension to effective dimensions by using his gale characterization of Hausdorff dimension [17]. Athreya, Hitchcock, Lutz, and Mayordomo then gave a dual gale characterization of packing dimension, with which, they generalized packing dimension to effective strong dimensions [1]. We first review the definitions related to gales. Note that $\Sigma_m$ is an alphabet with $m$ symbols and $m \geq 2$.

**Definition.** Let $s \in [0, \infty)$. An *s-supergale* is a function $d : \Sigma_m^* \to [0, \infty)$ such that for all $w \in \Sigma_m^*$

$$m^s d(w) \geq \sum_{a \in \Sigma_m} d(wa).$$

The *success set* of an *s*-supergale $d$ is

$$S^\infty[d] = \{S \in \mathbf{C} \mid \limsup_{n \to \infty} d(S[0..n-1]) = \infty\}.$$

The *strong success set* of $d$ is

$$S^\infty_{\text{str}}[d] = \{S \in \mathbf{C} \mid \liminf_{n \to \infty} d(S[0..n-1]) = \infty\}.$$

Now we conveniently give the gale characterizations of Hausdorff and packing dimensions as definitions. Please refer to Falconer [11] for classical definitions.

**Definition.** ([17, 1]). Let $X \subseteq \mathbf{C}_m$. The *Hausdorff dimension* of $X$ is

$$\dim_H(X) = \inf \{s \in [0, \infty) \mid X \subseteq S^\infty[d] \text{ for some } s\text{-supergale } d \}.$$

The *packing dimension* of $X$ is

$$\dim_P(X) = \inf \{s \in [0, \infty) \mid X \subseteq S^\infty_{\text{str}}[d] \text{ for some } s\text{-supergale } d\}.$$

Finite-state dimension and strong dimension are finite-state counterparts of classical Hausdorff dimension [15] and packing dimension [19, 27] introduced by Dai, Lathrop, Lutz, and Mayordomo [8] and Athreya, Hitchcock, Lutz, and Mayordomo [1] in the Cantor space $\mathbf{C}$. Finite-state dimensions are defined by using the gale characterizations of the Hausdorff dimension [17] and the packing dimension [1] and restricting the gales to the ones whose underlying betting strategies can be carried out by finite-state gamblers. In this section, we give the definitions of the finite-state dimensions for space $\mathbf{C}_m$ and review their basic properties. Now, we define finite-state gamblers on alphabet $\Sigma_m$.

**Definition.** ([8]) A *finite-state gambler* (*FSG*) is a 5-tuple $G = (Q, \Sigma_m, \delta, \vec{\beta}, q_0)$ such that $Q$ is a non-empty finite set of *states*; $\Sigma_m$ is the input alphabet; $\delta : Q \times \Sigma_m \to Q$ is the *state transition function*; $\vec{\beta} : Q \to \Delta(\Sigma_m)$ is the *betting function*; $q_0 \in Q$ is the *initial state*.

The extended transition function $\delta^* : Q \times \Sigma_m^* \to Q$ is defined such that

$$\delta^*(q, wa) = \begin{cases} q & \text{if } w = a = \lambda, \\ \delta(\delta^*(q, w), a) & \text{if } w \neq \lambda. \end{cases}$$

We use $\delta$ for $\delta^*$ and $\delta(w)$ for $\delta(q_0, w)$ for convenience.

The betting function $\beta_i : Q \to \Delta(\Sigma_m)$ specifies the bets the FSG places on each input symbol in $\Sigma_m$ with respect to a state $q \in Q$.

**Definition.** ([8]). Let $G = (Q, \Sigma_m, \delta, \vec{\beta}, q_0)$ be an FSG. The *s-gale* of $G$ is the function

$$d_G : \Sigma_m^* \to [0, \infty)$$

defined by the recursion

$$d_G(wb) = \begin{cases} 1 & \text{if } w = b = \lambda, \\ m^s d_G(w) \beta_i(\delta(w))(b) & \text{if } b \neq \lambda, \end{cases}$$

for all $w \in \Sigma_m^*$ and $b \in \Sigma_m \cup \{\lambda\}$. For $s \in [0, \infty)$, a function $d : \Sigma_m^* \to [0, \infty)$ is a *finite-state s-gale* if it is the $s$-gale of some finite-state gambler.

Note that in the original definition of a finite-state gambler the range of the betting function $\vec{\beta}$ is $\Delta(\{0, 1\}) \cap \mathbb{Q}^2$ [8, 1]. In the following observation, we show that allowing the range of $\vec{\beta}$ to have irrational probability measures does not change the notions of finite-state dimension and strong dimension.

**Observation 3.1.** Let $G = (Q, \Sigma_m, \delta, \vec{\beta}, q_0)$ be an FSG. For each $\epsilon > 0$, there exists an FSG $G = (Q, \Sigma_m, \delta, \vec{\beta}', q_0)$ with $\vec{\beta}' : Q \to \Delta(\Sigma_m) \cap \mathbb{Q}^m$ such that for all $s \in [0, \infty)$, $S^\infty[d_G^{(s)}] \subseteq S^\infty[d_{G'}^{(s+\epsilon)}]$ and $S_{\text{str}}^\infty[d_G^{(s)}] \subseteq S_{\text{str}}^\infty[d_{G'}^{(s+\epsilon)}]$.

In this paper, we allow the finite-state gamblers to place irrational bets.

**Definition.** ([8, 1]). Let $X \subseteq \mathbf{C}_m$. The *finite-state dimension* of $X$ is

$$\dim_{\text{FS}}(X) = \inf \{s \in [0, \infty) \mid X \subseteq S^\infty[d] \text{ for some finite-state } s\text{-gale } d\}$$

and the *finite-state strong dimension* of $X$ is

$$\text{Dim}_{\text{FS}}(X) = \inf \{s \in [0, \infty) \mid X \subseteq S_{\text{str}}^\infty[d] \text{ for some finite-state } s\text{-gale } d\}.$$

We will use the following basic properties of the Hausdorff, packing, finite-state, strong finite-state dimensions.

**Theorem 3.2.** *([8, 1]). Let $X, Y, X_i \subseteq \Sigma_m^\infty$ for $i \in \mathbb{N}$.*

1. $0 \leq \dim_H(X) \leq \dim_{\text{FS}}(X) \leq 1$, $0 \leq \dim_P(X) \leq \text{Dim}_{\text{FS}}(X) \leq 1$.
2. $\dim_H(X) \leq \dim_P(X)$, $\dim_{\text{FS}}(X) \leq \text{Dim}_{\text{FS}}(X)$.
3. *If $X \subseteq Y$, then the dimension of $X$ is at most that same dimension of $Y$.*
4. $\dim_{\text{FS}}(X \cup Y) = \max\{\dim_{\text{FS}}(X), \dim_{\text{FS}}(Y)\}$ *and* $\text{Dim}_{\text{FS}}(X \cup Y) = \max\{\text{Dim}_{\text{FS}}(X), \text{Dim}_{\text{FS}}(Y)\}$.
5. $\dim_H(\bigcup_{i=0}^\infty X_i) = \sup_{i \in \mathbb{N}} \dim_H(X_i)$, $\dim_P(\bigcup_{i=0}^\infty X_i) = \sup_{i \in \mathbb{N}} \dim_P(X_i)$.

# 4 Relative Frequencies of Digits

As we have mentioned in Section 1, Besicovitch in 1934 and Eggleston in 1949 proved the following two identities respectively.

**Theorem 4.1.** $\dim_H(\text{FREQ}^{\leq \beta}) = \mathcal{H}_2((\beta, 1 - \beta))$ *[3] and* $\dim_H(\text{FREQ}_\beta) = \mathcal{H}_2((\beta, 1 - \beta))$ *[9]*.

In this section, we will calculate the finite-state dimension of some more exotic sets that contain $m$-adic sequences that satisfy certain conditions placed on the frequencies of digits. The proofs in this section use straightforward constructions of finite-state gamblers. Both the constructions and analysis use completely elementary techniques.

Let $\mathcal{H}_{\beta,m}(\alpha) = -(\alpha \log_m \alpha + \beta\alpha \log_m \beta\alpha + (1 - \alpha - \beta\alpha) \log_m \frac{1 - \alpha - \beta\alpha}{m-2})$. Let

$$\alpha^*(x) = \begin{cases} \frac{1}{m} & x < 1 \\ \frac{1}{1 + x + (m-2)x^{\frac{x}{x+1}}} & \text{otherwise.} \end{cases}$$

Note that

$$\mathcal{H}_{\beta,m}(\alpha^*(\beta)) = \sup_{\alpha \in [0, \frac{1}{1+\beta}]} \mathcal{H}_{\beta,m}(\alpha) = \begin{cases} 1 & \text{if } \beta < 1, \\ \log_m(m - 2 + \frac{1+\beta}{\beta^{\frac{\beta}{\beta+1}}}) & \text{otherwise.} \end{cases}$$

**Theorem 4.2.** *Let $\beta' \geq \beta \geq 0$. Let*

$$X = \left\{ S \,\middle|\, \liminf_{n \to \infty} \frac{\pi_1(S, n)}{\pi_0(S, n)} \geq \beta \text{ and } \limsup_{n \to \infty} \frac{\pi_1(S, n)}{\pi_0(S, n)} \geq \beta' \right\}.$$

*Then $\dim_H(X) = \dim_{\text{FS}}(X) = \mathcal{H}_{\beta',m}(\alpha^*(\beta'))$ and $\dim_P(X) = \text{Dim}_{\text{FS}}(X) = \mathcal{H}_{\beta,m}(\alpha^*(\beta))$.*

**Corollary 4.3.** *(Theorem 2 [2]).* *Let $\beta \geq 0$. Let*

$$X = \left\{ S \,\middle|\, \lim_{n \to \infty} \frac{\pi_1(S,n)}{\pi_0(S,n)} = \beta \right\}.$$

*Let $\beta' = \max\{\beta, 1/\beta\}$. Then*

$$\dim_{\mathrm{H}}(X) = \mathcal{H}_{\beta,m}(\alpha^*(\beta')) = \log_m \left( m - 2 + \frac{1 + \beta'}{\beta'^{\frac{\beta'}{\beta'+1}}} \right)$$

Note that $\dim_{\mathrm{P}}(X)$, $\dim_{\mathrm{FS}}(X)$, and $\mathrm{Dim}_{\mathrm{FS}}(X)$ all takes the value of $\dim_{\mathrm{H}}(X)$, which were not proven in [2].

*Proof.* We prove the case where $\beta' = \beta$. The other case is similar. Let

$$Y = \left\{ S \,\middle|\, \liminf_{n \to \infty} \frac{\pi_1(S,n)}{\pi_0(S,n)} \geq \beta \right\}.$$

Let

$$Z = \left\{ S \,\middle|\, \begin{array}{l} \lim_{n \to \infty} \pi_0(S,n) = \alpha^*(\beta), \lim_{n \to \infty} \pi_1(S,n) = \beta\alpha^*(\beta), \\ \text{and } (\forall i > 1) \lim_{n \to \infty} \pi_i(S,n) = \frac{1 - \alpha^*(\beta) - \beta\alpha^*(\beta)}{m-2} \end{array} \right\}.$$

By Eggleston's theorem, $\dim_{\mathrm{H}}(Z) = \mathcal{H}_{\beta,m}(\alpha^*(\beta))$. Since $Z \subseteq X \subseteq Y$, it follows immediately from Theorem 4.2 that $\dim_{\mathrm{H}}(X) = \mathcal{H}_{\beta,m}(\alpha^*(\beta))$. □

## 5 Saturated Sets and Maximum Entropy Principle

In Section 4, we calculated the finite-state dimensions of many sets defined using properties on asymptotic frequencies of digits. They are all saturated sets. Now we formally define saturated sets and investigate their collective properties.

Let $\Pi_n(S) = \{\vec{\pi}(S,m) \mid m \geq n\}$ for all $n \in \mathbb{N}$. Let $\bar{\Pi}_n(S) = \overline{\Pi_n(S)}$, i.e., $\bar{\Pi}_n(S)$ is the closure of $\Pi_n(S)$. Define $\Pi : \mathbf{C}_m \to \mathcal{P}(\Delta(\Sigma_m))$ such that for all $S \in \mathbf{C}_m$, $\Pi(S) = \bigcap_{n \in \mathbb{N}} \bar{\Pi}_n(S)$.

**Definition.** Let $X \subseteq \mathbf{C}_m$. We say that *X is saturated* if for all $S, S' \in \mathbf{C}_m$,

$$\Pi(S) = \Pi(S') \Rightarrow [S \in X \iff S' \in X].$$

When we determine an upper bound on the finite-state dimensions of a set $X \subseteq \mathbf{C}_m$, it is in general not possible to use a single probability measure as the betting strategy even when $X$ is saturated. However, when certain conditions are true, a simple 1-state finite-state gambler may win on a huge set of sequences with different empirical digit distribution probability measures.

In the following, we formalize such a condition and reveal some relationship between betting and the Kullback-Leibler distance (relative entropy) [7]. Note that $m$-dimensional Kullback-Leibler distance $\mathcal{D}_m(\vec{\beta} \,\|\, \vec{\alpha})$ is defined as

$$\mathcal{D}_m(\vec{\beta} \,\|\, \vec{\alpha}) = \mathrm{E}_{\vec{\beta}} \log_m \frac{\vec{\beta}}{\vec{\alpha}}.$$

**Definition.** Let $\vec{\alpha}, \vec{\beta} \in \Delta(\Sigma_m)$. We say that $\vec{\alpha}$ *$\epsilon$-dominates* $\vec{\beta}$, denoted as $\vec{\alpha} \gg^\epsilon \vec{\beta}$, if $\mathcal{H}_m(\vec{\alpha}) \geq \mathcal{H}_m(\vec{\beta}) + \mathcal{D}_m(\vec{\beta} \,\|\, \vec{\alpha}) - \epsilon$. We say that $\vec{\alpha}$ *dominates* $\vec{\beta}$, denoted as $\vec{\alpha} \gg \vec{\beta}$, if $\vec{\alpha} \gg^0 \vec{\beta}$.

Note that $\mathcal{H}_m(\vec{\beta}) + \mathcal{D}_m(\vec{\beta} \,\|\, \vec{\alpha}) = \mathrm{E}_{\vec{\beta}} \log_m \frac{1}{\vec{\beta}} + \mathrm{E}_{\vec{\beta}} \log_m \frac{\vec{\beta}}{\vec{\alpha}} = \mathrm{E}_{\vec{\beta}} \log_m \frac{1}{\vec{\alpha}}$, where $\mathrm{E}_{\vec{\beta}} \log_m \frac{\vec{\beta}}{\vec{\alpha}} = \sum_{i=0}^{m-1} \beta_i \log_m \frac{\beta_i}{\alpha_i}$. It is very easy to see that the uniform probability measure dominates all probability measures.

**Observation 5.1.** *Let $\vec{\alpha} = (\frac{1}{m}, \ldots, \frac{1}{m})$. Let $\vec{\beta} \in \Delta(\Sigma_m)$. Then $\vec{\alpha} \gg \vec{\beta}$.*

Here, we give a few interesting properties of the domination relation.

**Theorem 5.2.** *Let $\vec{\alpha} = (\alpha_0, \ldots, \alpha_{k-1}) \in \Delta(\Sigma_k)$. Let $\vec{\beta} = (\beta_0, \ldots, \beta_{k-1}) \in \Delta(\Sigma_k)$ be such that $\beta_j = 1$, where $j = \arg\max\{\alpha_0, \ldots, \alpha_{k-1}\}$. Then $\vec{\alpha} >> \vec{\beta}$ and $\mathcal{H}_k(\vec{\beta}) = 0$.*

**Theorem 5.3.** *Let $\vec{\alpha}, \vec{\beta} \in \Delta(\Sigma_k)$, $\epsilon \geq 0$, and $r \in [0, 1]$. If $\vec{\alpha} >>^\epsilon \vec{\beta}$, then $\vec{\alpha} >>^\epsilon r\vec{\alpha} + (1-r)\vec{\beta}$.*

**Theorem 5.4.** *Let $\vec{\mu} = (\frac{1}{m}, \ldots, \frac{1}{m}) \in \Delta(\Sigma_m)$ be the uniform probability measure. Let $\vec{\beta} \in \Delta(\Sigma_m)$. Let $s \in [0,1]$. Let $\vec{\alpha} = s\vec{\mu} + (1-s)\vec{\beta}$. Then $\vec{\alpha} >> \vec{\beta}$.*

The following theorem relates the domination relation to finite-state dimensions.

**Theorem 5.5.** *Let $\vec{\alpha} \in \Delta(\Sigma_k)$ and $X \subseteq \Sigma_k^\infty$.*

1. *If $\vec{\alpha} >>^\epsilon \vec{\pi}(S, n)$ for infinitely many $n$ for every $\epsilon > 0$ and every $S \in X$, then $\dim_{\mathrm{FS}}(X) \leq \mathcal{H}_k(\vec{\alpha})$.*

2. *If $\vec{\alpha} >>^\epsilon \vec{\pi}(S, n)$ for all but finitely many $n$ for every $\epsilon > 0$ and every $S \in X$, then $\mathrm{Dim}_{\mathrm{FS}}(X) \leq \mathcal{H}_k(\vec{\alpha})$.*

Theorem 5.5 tells us that if we can find a single dominating probability measure for $X \subseteq \mathbf{C}_m$, then a simple 1-state FSG may be used to assess the dimension of $X$. However, in the following, we will see that the domination relationship is not even transitive.

**Theorem 5.6.** *Domination relation defined above is not transitive.*

Fix $\vec{\alpha} \in \Delta(\Sigma_m)$ with $\mathcal{H}_m(\vec{\alpha}) \neq 1$, the hyperplane $H$ in $\mathbb{R}^m$ defined by

$$\mathcal{H}_m(\vec{\alpha}) = \sum_{i=0}^{m-1} x_i \log_m \frac{1}{\alpha_i}$$

divides the simplex $\Delta(\Sigma_m)$ into two halves $A$ and $B$ with $A \cap B \subseteq H$. Suppose $(\frac{1}{m}, \ldots, \frac{1}{m}) \in B$, then $A = \{\vec{\beta} \in \Delta(\Sigma_m) \mid \vec{\alpha} >> \vec{\beta}\}$.

So it is not always possible to find a single probability measure that dominates all the empirical probability measures of sequences in $X \subseteq \mathbf{C}_m$. Nevertheless, we take advantage of the compactness of $\Delta(\Sigma_m)$ and give a general solution for finding the dimensions of $X \subseteq \mathbf{C}_m$, when $X$ is saturated. The following theorem is our pointwise maximum entropy principle for saturated sets. It says that the dimension of a saturated set is the maximum pointwise asymptotic entropy of the empirical digit distribution measure.

**Theorem 5.7.** *Let $X \subseteq \mathbf{C}_m$ be saturated. Let*

$$H = \sup_{S \in X} \liminf_{n \to \infty} \mathcal{H}_m(\vec{\pi}(S, n))$$

*and*

$$P = \sup_{S \in X} \limsup_{n \to \infty} \mathcal{H}_m(\vec{\pi}(S, n)).$$

*Then $\dim_{\mathrm{FS}}(X) = \dim_{\mathrm{H}}(X) = H$ and $\mathrm{Dim}_{\mathrm{FS}}(X) = \dim_{\mathrm{P}}(X) = P$.*

This theorem automatically gives a solution for finding an upper bounds for dimensions of arbitrary $X$.

**Corollary 5.8.** *Let $X \subseteq \mathbf{C}_m$ and let $H$ and $P$ be defined as in Theorem 5.7. Then $\dim_{\mathrm{FS}}(X) \leq H$ and $\mathrm{Dim}_{\mathrm{FS}}(X) \leq P$.*

In the following, we derive the dimensions of a few interesting saturated sets using Theorem 5.7. We will give more examples in the full version of this paper.

Let $H_{\alpha, m} = \log_m[\alpha^{-\alpha}(\frac{1-\alpha}{m-1})^{\alpha-1}]$.

**Theorem 5.9.** *Let $\underline{\alpha}, \bar{\alpha} \in [0,1]$ such that $1/m < \underline{\alpha} \le \bar{\alpha}$ and let*
$$M_k^{\underline{\alpha},\bar{\alpha}} = \{S \in \Sigma_m^\infty \mid \liminf_{n\to\infty} \pi_k(S,n) = \underline{\alpha} \text{ and } \limsup_{n\to\infty} \pi_k(S,n) = \bar{\alpha}\}.$$

*Then $\dim_{\mathrm{H}}(M_k^{\underline{\alpha},\bar{\alpha}}) = H_{\bar{\alpha},m}$ and $\dim_{\mathrm{P}}(M_k^{\underline{\alpha},\bar{\alpha}}) = H_{\underline{\alpha},m}$.*

*Proof.* It is easy to check that $M_k^{\underline{\alpha},\bar{\alpha}}$ is saturated, that $H_{\bar{\alpha},m} = \inf_{\alpha \in [\underline{\alpha},\bar{\alpha}]} H_{\alpha,m}$, and that $H_{\underline{\alpha},m} = \sup_{\alpha \in [\underline{\alpha},\bar{\alpha}]} H_{\alpha,m}$. The theorem follows from Theorem 5.7. □

**Corollary 5.10.** *Let $\underline{\alpha}, \bar{\alpha} \in [0,1]$ such that $\underline{\alpha} \le \bar{\alpha}$ and let*
$$M_k^{\underline{\alpha},\bar{\alpha}} = \{S \in \mathbf{C}_m \mid \liminf_{n\to\infty} \pi_k(S,n) = \underline{\alpha} \text{ and } \limsup_{n\to\infty} \pi_k(S,n) = \bar{\alpha}\}.$$

*Then*
$$\dim_{\mathrm{H}}(M_k^{\underline{\alpha},\bar{\alpha}}) = \inf_{\alpha \in [\underline{\alpha},\bar{\alpha}]} H_{\alpha,m} = \min(H_{\underline{\alpha},m}, H_{\bar{\alpha},m})$$

*and*
$$\dim_{\mathrm{P}}(M_k^{\underline{\alpha},\bar{\alpha}}) = \sup_{\alpha \in [\underline{\alpha},\bar{\alpha}]} H_{\alpha,m} = \begin{cases} 1 & \underline{\alpha} \le 1/m \le \bar{\alpha}, \\ \max(H_{\underline{\alpha},m}, H_{\bar{\alpha},m}) & \text{otherwise.} \end{cases}$$

*Proof.* If $\underline{\alpha} \le 1/m \le \bar{\alpha}$, then for some $S \in M_k^{\underline{\alpha},\bar{\alpha}}$, $\limsup_{n\to\infty} \mathcal{H}_m(\vec{\pi}(S,n)) = 1$. □

**Corollary 5.11.** *(Theorem 7 [2]). Let $\underline{\alpha}_k, \bar{\alpha}_k \in [0,1]$ for $k \in \Sigma_m$. Let $M_R = \bigcap_{k=0}^{m-1} M_k^{\underline{\alpha}_k,\bar{\alpha}_k}$. Then*
$$\dim_{\mathrm{FS}}(M_R) = \dim_{\mathrm{H}}(M_R) = \min_{k=0}^{m-1} \dim_{\mathrm{H}}(M_k^{\underline{\alpha},\bar{\alpha}})$$

*and*
$$\mathrm{Dim}_{\mathrm{FS}}(M_R) = \dim_{\mathrm{P}}(M_R) = \min_{k=0}^{m-1} \dim_{\mathrm{P}}(M_k^{\underline{\alpha},\bar{\alpha}}).$$

**Corollary 5.12.** *(Theorem 1 [2]). Let $k \in \Sigma_m$ and let*
$$M_k = \{S \in \mathbf{C}_m \mid \liminf_{n\to\infty} \pi_k(x,n) < \limsup_{n\to\infty} \pi_k(x,n)\}.$$

*Then $\dim_{\mathrm{H}}(\bigcap_{k=0}^{m-1} M_k) = 1$.*

**Theorem 5.13.** *Let $A$ be a $d \times m$ matrix and $b = (b_1, \ldots, b_d) \in \mathbb{R}^d$. Let*
$$K^{\text{i.o.}}(A,b) = \{S \in \mathbf{C}_m \mid (\exists \{k_n\} \subseteq \mathbb{N}) \lim_{n\to\infty} k_n = \infty \text{ and } \lim_{n\to\infty} A(\vec{\pi}(S,k_n))^T = b\}$$

*and let $K(A,b) = \{S \in \mathbf{C}_m \mid \lim_{n\to\infty} A(\vec{\pi}(S,n))^T = b\}$. Then*
$$\dim_{\mathrm{FS}}(K^{\text{i.o.}}(A,b)) = \dim_{\mathrm{H}}(K^{\text{i.o.}}(A,b)) = \sup_{\substack{\vec{\alpha} \in \Delta(\Sigma_m) \\ A\vec{\alpha}^T = b}} \mathcal{H}_m(\vec{\alpha}),$$

$\dim_{\mathrm{P}}(K^{\text{i.o.}}(A,b)) = 1$, *and*
$$\dim_{\mathrm{H}}(K(A,b)) = \mathrm{Dim}_{\mathrm{FS}}(K(A,b)) = \sup_{\substack{\vec{\alpha} \in \Delta(\Sigma_m) \\ A\vec{\alpha}^T = b}} \mathcal{H}_m(\vec{\alpha}).$$

*Proof.* It is easy to check that $K^{\text{i.o.}}(A,b)$ and $K(A,b)$ are both saturated. □

# 6 Conclusion

A general saturated set usually has an uncountable decomposition in which, the dimension of each element is easy to determine, while the dimension of the whole set, which is the uncountable union of all the element sets, is very difficult to determine and requires advanced techniques in multifractal analysis and ergodic theory. By using finite-state gambler and gale characterizations of dimensions, we are able to obtain very general results calculating the classical dimensions and finite-state dimensions of saturated sets using completely elementary analysis. This indicates that gale characterizations will play a more important role in dimension-theoretic analysis and that finite-state gambler is very powerful.

# A  Appendix for Section 3

**Proof of Observation 3.1.** Let $\delta = \min_{q \in Q} \min_{i=0..m}(\beta_i(q) - \frac{\beta_i(q)}{2^\epsilon})$. For all $q \in Q$, let $\beta'_i(q) \in [\beta_i(q) - \delta, \beta_i(q)] \cap [0,1] \cap \mathbb{Q}$, if $i \neq 0$. Otherwise, let $\beta'_0(q) = 1 - \sum_{i=1}^{m-1} \beta'_i(q)$. Note that for all $q \in Q$ and all $i \in \{0, \ldots, m-1\}$,
$$\beta'_i(q) \geq \beta_i(q) - \delta \geq 0 \tag{A.1}$$
and that $\beta'$ maps states to rational bets.

For $w \in \Sigma_m^*$, let
$$\#q_i(w) = |\{n \mid 0 \leq n \leq |w| - 1, \delta(w[0..n-1]) = q \text{ and } w[n] = i\}|.$$

Note that
$$|w| = \sum_{q \in Q} \sum_{i=0}^{m-1} \#q_i(w). \tag{A.2}$$

Now we have that for all $w \in \Sigma_m^*$,

$$d_{G'}^{(s+\epsilon)}(w) = c_0 m^{(s+\epsilon)|w|} \prod_{q \in Q} \prod_{i=0}^{m-1} \beta'_i(q)^{\#q_i(w)}$$

$$\geq^{(A.1)} c_0 m^{(s+\epsilon)|w|} \prod_{q \in Q} \prod_{i=0}^{m-1} (\beta_i(q) - \delta)^{\#q_i(w)}$$

$$=^{(A.2)} c_0 m^{s|w|} \prod_{q \in Q} \prod_{i=0}^{m-1} [(\beta_i(q) - \delta)2^\epsilon]^{\#q_i(w)}.$$

By the choice of $\delta$,

$$d_{G'}^{(s+\epsilon)}(w) \geq c_0 m^{s|w|} \prod_{q \in Q} \prod_{i=0}^{m-1} [(\beta_i(q) - (\beta_i(q) - \frac{\beta_i(q)}{2^\epsilon}))2^\epsilon]^{\#q_i(w)}$$

$$= c_0 m^{s|w|} \prod_{q \in Q} \prod_{i=0}^{m-1} \beta_i(q)^{\#q_i(w)}$$

$$= d_G^{(s)}(w).$$

Thus, for all $S \in S^\infty[d_G^{(s)}]$,
$$\limsup_{n \to \infty} d_{G'}^{(s+\epsilon)}(S) \geq \limsup_{n \to \infty} d_G^{(s)}(S) = \infty,$$

and for all $S \in S_{\text{str}}^\infty[d_G^{(s)}]$,
$$\liminf_{n \to \infty} d_{G'}^{(s+\epsilon)}(S) \geq \liminf_{n \to \infty} d_G^{(s)}(S) = \infty.$$

Therefore,
$$S^\infty[d_G^{(s)}] \subseteq S^\infty[d_{G'}^{(s+\epsilon)}]$$
and
$$S_{\text{str}}^\infty[d_G^{(s)}] \subseteq S_{\text{str}}^\infty[d_{G'}^{(s+\epsilon)}].$$

□

# B  Appendix for Section 4

**Proof of Theorem 4.2.** We assume that $\beta' \geq \beta \geq 1$, since when either of these values are less than 1, the proof is essentially looking at the subset of $X$ where their values are replaced by 1. First, we prove the lower bounds for the dimensions.

When $S$ is clear from the context, let $\alpha_n = \pi_0(S, n)$ and $\beta_n = \pi_1(S, n)$.

Let $\alpha' = \alpha^*(\beta')$ and let $\alpha = \alpha^*(\beta)$.

For Hausdorff dimension and finite-state dimension, let

$$Y = \left\{ S \ \middle| \ \lim_{n \to \infty} \alpha_n = \alpha', \lim_{n \to \infty} \beta_n = \beta'\alpha', \text{and } (\forall i > 1) \lim_{n \to \infty} \pi_i(S, n) = \frac{1 - \alpha' - \beta'\alpha'}{m - 2} \right\}.$$

By Eggleston's theorem, we have $\dim_H(Y) = \mathcal{H}_{\beta',m}(\alpha^*(\beta'))$. Since $\beta' \geq \beta \geq 1$ and $Y \subseteq X$,

$$\dim_{FS}(X) \geq \dim_H(X) \geq \dim_H(Y) = \mathcal{H}_{\beta',m}(\alpha^*(\beta')).$$

For packing dimension and finite-state strong dimension, let

$$Z = \left\{ S \ \middle| \ \lim_{n \to \infty} \alpha_n = \alpha, \lim_{n \to \infty} \beta_n = \beta\alpha, \text{and } (\forall i > 1) \lim_{n \to \infty} \pi_i(S, n) = \frac{1 - \alpha - \beta\alpha}{m - 2} \right\}.$$

Now we construct from $Z$ a set $Z' \subseteq X$ by interpolating the sequences in $Z$.

First let $l_0 = 2$ and for every $i \in \mathbb{N}$, $l_{i+1} = 2^{l_i}$.

Define $f_0 : \Sigma_m^* \to \Sigma_m^*$ be such that $f_0(w) = w$ for all $w \in \Sigma_m^*$. Let $\rho = \frac{1}{\alpha\beta' - \alpha\beta + 1}$. For each $n > 0$, define $f_n : \Sigma_m^* \to \Sigma_m^*$ such that for every $w \in \Sigma_m^*$, $|f_n(w)| = |w|$ and for every $i < |w|$,

$$f_n(w)[i] = \begin{cases} f_{n-1}(w)[i] & i \leq l_{n-1} \\ w[i] & i \leq \lceil \rho l_n \rceil \text{ and } i > l_{n-1} \\ 1 & i > \lceil \rho l_n \rceil \text{ and } i \leq l_n \\ w[i] & i > l_n. \end{cases}$$

Define $f : \Sigma_m^* \to \Sigma_m^*$ such that for all $w \in \Sigma_m^*$

$$f(w) = f_{n(w)}(w),$$

where $n(w) = \min \{ n \in \mathbb{N} \mid l_n \geq |w| \}$. Also, extend $f$ to $f : \Sigma_m^\infty \to \Sigma_m^\infty$ such that for all $S \in \Sigma_m^\infty$,

$$f(S) = \lim_{n \to \infty} f(S[0..n-1]).$$

Let $Z' = f(Z)$.

By the construction of $f$ and choice of $\rho$, it is clear that $f$ is a dilation and for all $n \in \mathbb{N}$, $|\mathrm{Col}(f, S[0.. \lceil \rho l_n \rceil - 1])| \leq \log l_n$. Thus for all $\epsilon > 0$, there are infinitely many $n$ such that

$$|\mathrm{Col}(f, S[0..n-1])| < \epsilon n. \tag{B.1}$$

Note that by Eggleston's theorem, $\dim_H(Z) = \mathcal{H}_{\beta,m}(\alpha^*(\beta))$. Then by Supergale Dilation Theorem [13] and (B.1), $\dim_P(Z') \geq \mathcal{H}_{\beta,m}(\alpha^*(\beta))$.

It is easy to verify that for every $S \in Z'$,

$$\liminf_{n \to \infty} \frac{\beta_n}{\alpha_n} \geq \beta \text{ and } \limsup_{n \to \infty} \frac{\beta_n}{\alpha_n} \geq \beta'.$$

So $Z' \subseteq X$. Therefore, $\mathrm{Dim}_{FS}(X) \geq \dim_P(X) \geq \mathcal{H}_{\beta,m}(\alpha^*(\beta))$.

Now, we prove that $\mathcal{H}_{\beta',m}(\alpha^*(\beta'))$ is an upper bound for $\dim_H(X)$ and $\dim_{FS}(X)$.

If $\beta' < 1$, then $\mathcal{H}_{\beta',m}(\alpha^*(\beta')) = 1$ and the upper bound holds trivially. So assume $\beta' \geq 1$.

Let $\alpha = \alpha^*(\beta')$. Let $s > \mathcal{H}_{\beta',m}(\alpha^*(\beta'))$. Define

$$d(wb) = \begin{cases} m^s \alpha d(w) & b = 0 \\ m^s \beta' \alpha d(w) & b = 1 \\ m^s \frac{1-\alpha-\beta'\alpha}{m-2} d(w) & b \geq 2 \end{cases}.$$

It is clear that $d$ is a finite-state $s$-gale.

Let
$$B = \beta'^{\frac{\beta'}{\beta'+1}}.$$

Let
$$\epsilon = \frac{s - \mathcal{H}_{\beta',m}(\alpha^*(\beta'))}{2 \log_m B}.$$

Let $S \in X$ and let $\delta > 0$ be such that $\delta \leq \min(\epsilon \beta'^2/2, 1/2)$.

Since
$$\limsup_{n \to \infty} \frac{\beta_n}{\alpha_n} \geq \beta',$$

there exists an infinite set $J \subseteq \mathbb{N}$ such that for all $n \in J$

$$\frac{\beta_n}{\alpha_n} \geq \beta' - \delta.$$

By the choice of $\delta$, for all $n \in J$

$$\frac{\alpha_n}{\beta_n} \leq \frac{1}{\beta'-\delta} = \frac{1}{\beta'} + \frac{\delta}{(\beta'-\delta)\beta'} \leq \frac{1}{\beta'} + \epsilon,$$

i.e.,
$$\alpha_n + \beta_n \leq \frac{\beta'+1}{\beta'} \beta_n + \epsilon. \tag{B.2}$$

Now, note that
$$m^s B^{1-\epsilon} = (1 + \beta' + (m-2)B)B^\epsilon, \tag{B.3}$$

since

$$\begin{aligned}
m^s B^{1-\epsilon} &= m^s B^{1-\frac{s-\log_m(m-2+\frac{1+\beta'}{B})}{2 \log_m B}} \\
&= B^{1+\log_B m^s - \frac{\log_m m^s - \log_m(m-2+\frac{1+\beta'}{B})}{2 \log_m B}} \\
&= B^{1+\frac{2\log_m m^s - \log_m m^s + \log_m(m-2+\frac{1+\beta'}{B})}{2 \log_m B}} \\
&= B^{1+\frac{\log_m m^s + \log_m(m-2+\frac{1+\beta'}{B})}{2 \log_m B}} \\
&= B^{1+\frac{s-\log_m(m-2+\frac{1+\beta'}{B})+2\log_m(m-2+\frac{1+\beta'}{B})}{2 \log_m B}} \\
&= B^{1+\epsilon+\log_B(m-2+\frac{1+\beta'}{B})}.
\end{aligned}$$

For all $n \in J$,

$$d(S[0..n-1]) = m^{sn}\alpha^{n\alpha_n}(\beta'\alpha)^{n\beta_n}\left(\frac{1-\alpha-\beta'\alpha}{m-2}\right)^{n(1-\alpha_n-\beta_n)}$$

$$= \left[\frac{m^s\beta'^{\beta_n}B^{1-\alpha_n-\beta_n}}{1+\beta'+(m-2)B}\right]^n$$

$$\geq^{(B.2)} \left[\frac{m^s\beta'^{\beta_n}B^{1-\frac{\beta'+1}{\beta'}\beta_n-\epsilon}}{1+\beta'+(m-2)B}\right]^n$$

$$= \left[\frac{m^s B^{1-\epsilon}}{1+\beta'+(m-2)B}\right]^n$$

$$=^{(B.3)} B^{\epsilon n}.$$

Since $J$ is an infinite set,
$$\limsup_{n\to\infty} d(S[0..n-1]) = \infty,$$
i.e., $S \in S^\infty[d]$. Since $s > \mathcal{H}_{\beta',m}(\alpha^*(\beta'))$ is arbitrary and $d$ is finite-state $s$-gale, $\dim_H(X) \leq \dim_{FS}(X) \leq \mathcal{H}_{\beta',m}(\alpha^*(\beta'))$.

An essentially identical argument gives us $\dim_P(X) \leq \text{Dim}_{FS}(X) \leq \mathcal{H}_{\beta,m}(\alpha^*(\beta))$. □

**Theorem B.1.** Let $\alpha \geq 1/m$. Let $X = \left\{S \mid \lim_{n\to\infty} \pi_0(S,n) = \alpha\right\}$ and $Y = \left\{S \mid \liminf_{n\to\infty} \pi_0(S,n) \geq \alpha\right\}$. Then
$$\dim_P(X) = \dim_H(X) = \dim_P(Y) = \dim_H(Y) = \log_m\left[\alpha^{-\alpha}\left(\frac{1-\alpha}{m-1}\right)^{\alpha-1}\right].$$

**Proof of Theorem B.1.** The results are clear for $\alpha = 1/m$, so we assume that $\alpha > 1/m$.

Let $H_{\alpha,m} = \log_m\left[\alpha^{-\alpha}\left(\frac{1-\alpha}{m-1}\right)^{\alpha-1}\right]$.

We first show that $\dim_P(Y) \leq H_{\alpha,m}$. For $s > H_{\alpha,m}$, define

$$d(wb) = \begin{cases} m^s\alpha d(w) & b = 0 \\ m^s\frac{1-\alpha}{m-1}d(w) & b \neq 0. \end{cases}$$

It is clear that $d$ is an $s$-gale. Let

$$\epsilon = \frac{s - H_{\alpha,m}}{2\log_m\frac{\alpha(m-1)}{1-\alpha}}. \tag{B.4}$$

Note that $\frac{\alpha(m-1)}{1-\alpha} > 1$. Let $S \in Y$, i.e., $\liminf_{n\to\infty} \pi_0(S,n) \geq \alpha$. So there exists $J \subseteq \mathbb{N}$ such that $|\mathbb{N} \setminus J| < \infty$ and for every $n \in J$,

$$\pi_0(S,n) \geq \alpha - \epsilon.$$

$$d(S[0..n-1]) = \left[m^s \alpha^{\pi_0(S,n)} \left(\frac{1-\alpha}{m-1}\right)^{1-\pi_0(S,n)}\right]^n$$

$$\stackrel{\text{(B.4)}}{=} \left[\left(\frac{\alpha(m-1)}{1-\alpha}\right)^{2\epsilon} \alpha^{-\alpha} \left(\frac{1-\alpha}{m-1}\right)^{\alpha-1} \alpha^{\pi_0(S,n)} \left(\frac{1-\alpha}{m-1}\right)^{1-\pi_0(S,n)}\right]^n$$

$$= \left[\left(\frac{\alpha(m-1)}{1-\alpha}\right)^{2\epsilon} \alpha^{\pi_0(S,n)-\alpha} \left(\frac{1-\alpha}{m-1}\right)^{\alpha-\pi_0(S,n)}\right]^n$$

$$= \left[\left(\frac{\alpha(m-1)}{1-\alpha}\right)^{2\epsilon} \left(\frac{\alpha(m-1)}{1-\alpha}\right)^{\pi_0(S,n)-\alpha}\right]^n$$

$$= \left[\left(\frac{\alpha(m-1)}{1-\alpha}\right)^{2\epsilon+\pi_0(S,n)-\alpha}\right]^n.$$

Then for every $n \in J$,
$$d(S[0..n-1]) \geq \left[\frac{\alpha(m-1)}{1-\alpha}\right]^{\epsilon n}.$$

Since $\frac{\alpha(m-1)}{1-\alpha} > 1$, $S \in S^\infty_{\text{str}}[d]$ and $\dim_H(Y) \leq \dim_P(Y) \leq H_{\alpha,m}$. Note taht $X \subseteq Y$, so $\dim_H(X) \leq \dim_P(X) \leq H_{\alpha,m}$.

Now it suffices to show that $\dim_H(X) \geq H_{\alpha,m}$.

Let
$$Z = \left\{S \,\middle|\, \lim_{n\to\infty} \pi_0(S[0..n-1]) = \alpha \text{ and } (\forall i > 0) \lim_{n\to\infty} \pi_i(S[0..n-1]) = \frac{1-\alpha}{m-1}\right\}.$$

By Eggleston's theorem, $\dim_H(Z) = H_{\alpha,m}$. Since $Z \subseteq X \subseteq Y$, $\dim_H(Y) \geq \dim_H(X) \geq H_{\alpha,m}$. □

**Theorem B.2.** *(Corollary 13 in [2]). Let $\Sigma_m$ be the m-ary alphabet. Let $k < m$. Let $\alpha_0, \alpha_1, \ldots, \alpha_{k-1} \in [0,1]$ be such that $\alpha = \sum_{i=0}^{k-1} \alpha_i \leq 1$. Let*
$$X = \left\{S \,\middle|\, \lim_{n\to\infty} \pi_i(S,n) = \alpha_i, 0 \leq i \leq k\right\}.$$

*Then $\dim_H(X)$ is*
$$\mathcal{H}_m\left(\alpha_0, \ldots, \alpha_{k-1}, \tfrac{1-\alpha}{m-k}, \ldots, \tfrac{1-\alpha}{m-k}\right) = \log_m\left[\alpha_0^{-\alpha_0} \cdots \alpha_{k-1}^{-\alpha_{k-1}} \left(\tfrac{1-\alpha}{m-k}\right)^{-(1-\alpha)}\right]$$

*and*
$$\dim_{\text{FS}}(X) = \text{Dim}_{\text{FS}}(X) = \dim_P(X) = \dim_H(X).$$

**Proof of Theorem B.2.** We insist that $0^0 = 1$ and $0/0 = 1$ in the proof.

Let
$$H = \mathcal{H}_m\left(\alpha_0, \alpha_1, \ldots, \alpha_{k-1}, \frac{1-\alpha}{m-k}, \ldots, \frac{1-\alpha}{m-k}\right).$$

For $s > H$, define
$$d(wb) = \begin{cases} m^s d(w) \alpha_b & b < k \\ m^s d(w) \frac{1-\alpha}{m-k} & \text{otherwise.} \end{cases}$$

It is clear that $d$ is a finite-state $s$-gale. Let
$$\delta = \frac{s-H}{-2\log_m(\alpha_0 \cdots \alpha_{k-1} \frac{1-\alpha}{m-k})}.$$

For $S \in X$,
$$\lim_{n\to\infty} \pi_i(S,n) = \alpha_i, 0 \leq i \leq k.$$

So there exists $n_0 \in \mathbb{N}$ such that for all $n \geq n_0$ $|\pi_i(S,n) - \alpha_i| < \delta$ for all $i < k$ and that

$$\left| \alpha - \sum_{i=0}^{k-1} \pi_i(S,n) \right| < \delta$$

Then for all $n \geq n_0$,

$$d(S[0..n-1]) = \left[ m^s \left( \frac{1-\alpha}{m-k} \right)^{1-\sum_{i=0}^{k-1} \pi_i(S,n)} \prod_{i=0}^{k-1} \alpha_i^{\pi_i(S,n)} \right]^n$$

$$= \left[ m^{s-H} m^H \left( \frac{1-\alpha}{m-k} \right)^{1-\sum_{i=0}^{k-1} \pi_i(S,n)} \prod_{i=0}^{k-1} \alpha_i^{\pi_i(S,n)} \right]^n$$

$$= \left[ m^{s-H} \alpha_0^{-\alpha_0} \cdots \alpha_{k-1}^{-\alpha_{k-1}} \left( \frac{1-\alpha}{m-k} \right)^{-(1-\alpha)} \left( \frac{1-\alpha}{m-k} \right)^{1-\sum_{i=0}^{k-1} \pi_i(S,n)} \prod_{i=0}^{k-1} \alpha_i^{\pi_i(S,n)} \right]^n$$

$$= \left[ m^{s-H} \left( \frac{1-\alpha}{m-k} \right)^{\alpha - \sum_{i=0}^{k-1} \pi_i(S,n)} \prod_{i=0}^{k-1} \alpha_i^{\pi_i(S,n) - \alpha_i} \right]^n$$

$$\geq \left[ m^{s-H} \left( \alpha_0 \cdots \alpha_{k-1} \frac{1-\alpha}{m-k} \right)^{\delta} \right]^n = \left[ m^{s-H} m^{\frac{H-s}{2}} \right]^n$$

$$= m^{\frac{s-H}{2} n}.$$

So $S \in S_{\text{str}}^{\infty}[d]$ and thus $\dim_{\text{FS}}(X) \leq \text{Dim}_{\text{FS}}(X) \leq H$.

Let

$$Z = \left\{ S \,\bigg|\, (\forall i < k) \lim_{n \to \infty} \pi_i(S,n) = \alpha_i \text{ and } (\forall i \geq k) \lim_{n \to \infty} \pi_i(S,n) = \frac{1-\alpha}{m-k} \right\}.$$

By Eggleston's theorem, $\dim_{\text{H}}(Z) = H$. The theorem then follows from the monotonicity of dimensions. $\square$

## C   Appendix for Section 5

**Proof of Theorem 5.2.** It is easy to see that $\mathcal{H}_k(\beta) = 0$. It suffices to show that

$$\mathcal{H}_k(\vec{\alpha}) \geq \mathrm{E}_{\vec{\beta}} \log_k \frac{1}{\vec{\alpha}}.$$

$$\mathrm{E}_{\vec{\beta}} \log_k \frac{1}{\vec{\alpha}} = \sum_{i=0}^{k-1} \beta_i \log_k \frac{1}{\alpha_i} = \beta_j \log_k \frac{1}{\alpha_j}$$

$$= \log_k \frac{1}{\alpha_j} \leq \sum_{i=0}^{k-1} \alpha_i \log_k \frac{1}{\alpha_i}$$

$$= \mathcal{H}_k(\vec{\alpha}).$$

$\square$

**Proof of Theorem 5.3.** Assume $\vec{\alpha} >>^\epsilon \vec{\beta}$, it suffices to show that

$$\mathcal{H}_k(\vec{\alpha}) \geq \mathrm{E}_{r\vec{\alpha}+(1-r)\vec{\beta}} \log_k \frac{1}{\vec{\alpha}} - \epsilon.$$

$$\mathrm{E}_{r\vec{\alpha}+(1-r)\vec{\beta}} \log_k \frac{1}{\vec{\alpha}} - \epsilon = \sum_{i=0}^{k-1}(r\alpha_i + (1-r)\beta_i)\log_k \frac{1}{\alpha_i} - \epsilon$$

$$= \sum_{i=0}^{k-1} r\alpha_i \log_k \frac{1}{\alpha_i} + \sum_{i=0}^{k-1}(1-r)\beta_i \log_k \frac{1}{\alpha_i} - \epsilon$$

$$= r\mathcal{H}_k(\vec{\alpha}) + (1-r)\mathrm{E}_{\vec{\beta}}\log_k \frac{1}{\vec{\alpha}} - (1-r)\epsilon - r\epsilon$$

$$\leq \mathcal{H}_k(\vec{\alpha}).$$

$\square$

**Proof of Theorem 5.4.** Let $A = \{i \mid \mu_i \geq \beta_i\}$ and let $B = \{i \mid \mu_i < \beta_i\}$. Then $A \cap B = \varnothing$ and $A \cup B = [0..m-1]$. Note that for any $i \in A$, $\mu_i = \frac{1}{m} \geq \beta_i$ and $\log_m \frac{1}{s\mu_i+(1-s)\beta_i} \geq 1$ and for any $i \in B$, $\mu_i = \frac{1}{m} < \beta_i$ and $\sum_{i \in B} s(\mu_i - \beta_i)\log_m \frac{1}{s\mu_i+(1-s)\beta_i} < 1$. Since $\sum_{i=0}^{m-1} s(\mu_i - \beta_i) = 0$, $\sum_{i \in A} s(\mu_i - \beta_i) = -\sum_{i \in B} s(\mu_i - \beta_i)$.

$$\mathrm{E}_{\vec{\alpha}} \log_m \frac{1}{\vec{\alpha}} - \mathrm{E}_{\vec{\beta}}\log_m \frac{1}{\vec{\alpha}}$$

$$= \mathrm{E}_{s(\vec{\mu}-\vec{\beta})} \log_m \frac{1}{s\vec{\mu}+(1-s)\vec{\beta}}$$

$$= \sum_{i=0}^{m-1} s(\mu_i - \beta_i) \log_m \frac{1}{s\mu_i + (1-s)\beta_i}$$

$$= \sum_{i \in A} s(\mu_i - \beta_i) \log_m \frac{1}{s\mu_i + (1-s)\beta_i} + \sum_{i \in B} s(\mu_i - \beta_i) \log_m \frac{1}{s\mu_i + (1-s)\beta_i}$$

$$\geq \sum_{i \in A} s(\mu_i - \beta_i) \cdot 1 + \sum_{i \in B} s(\mu_i - \beta_i) \cdot 1$$

$$\geq 0.$$

Therefore,
$$\mathrm{E}_{\vec{\alpha}} \log_m \frac{1}{\vec{\alpha}} \geq \mathrm{E}_{\vec{\beta}} \log_m \frac{1}{\vec{\alpha}},$$
i.e., $\vec{\alpha} >> \vec{\beta}$.

$\square$

Proof of Theorem 5.5. Let $G = (Q, \delta, \vec{\beta}, q_0, 1)$ be an FSG such that $Q = \{q_0\}$, $\delta(q_0, b) = q_0$ for all $b \in \Sigma_k$, and $\vec{\beta}(q_0) = \vec{\alpha}$.

Let $s > \mathcal{H}_k(\vec{\alpha}) + \epsilon$. The $s$-gale $d_G^{(s)}$ of $G$ is defined by the following recursion,

$$d_G^{(s)}(wb) = \begin{cases} 1 & w = b = \lambda \\ k^s d_G^{(s)}(w)\alpha_b & \text{otherwise,} \end{cases}$$

for all $w \in \Sigma_k^*$ and $b \in \Sigma_k$. Let $S \in X$. Then

$$d_G^{(s)}(S[0..n-1]) = k^{sn} \prod_{i=0}^{k-1} \alpha_i^{n\pi_i(S,n)}$$

$$= k^{sn} k^{n \sum_{i=0}^{k-1} \pi_i(S,n) \log_k \alpha_i}$$

$$= \left(k^{s - \mathrm{E}_{\vec{\pi}(S,n)} \log_k \frac{1}{\vec{\alpha}}}\right)^n.$$

Thus $S \in S^\infty[d_G^{(s)}]$ and $\dim_{\mathrm{FS}}(S) \leq s$, when the domination condition holds for infinitely many $n$. Similarly, $S \in S^\infty_{\mathrm{str}}[d_G^{(s)}]$ and $\mathrm{Dim}_{\mathrm{FS}}(S) \leq s$, when the domination condition holds for all but finitely many $n$. The theorem then follows, since $\epsilon$ can be arbitrarily small.

$\square$

**Proof of Theorem 5.6.** We prove this by giving a counterexample with $\Sigma_3$. This counterexample can be extended to larger alphabets very easily.

Let $\alpha = (\frac{54}{300}, \frac{54}{300}, \frac{192}{300})$, $\beta = (\frac{25}{300}, \frac{75}{300}, \frac{200}{300})$. And we have

$$\mathcal{H}(\alpha) \approx 0.8219015831,$$

and

$$\mathcal{H}(\beta) \approx 0.7344147903,$$

and

$$\mathcal{H}(\alpha) - \mathrm{E}_\beta \log_3 \frac{\beta}{\alpha} \approx 0.05003477990.$$

So $\alpha >> \beta$.

Note that fix $\alpha \in \Delta(\{0, 1, 2\})$, for $\gamma \in \Delta(\{0, 1, 2\})$, $\alpha >> \gamma$ if

$$\mathcal{H}(\alpha) \geq \mathrm{E}_\gamma \log_3 \frac{1}{\alpha},$$

i.e.,

$$\mathcal{H}(\alpha) \geq \gamma_0 \log_3 \frac{1}{\alpha_0} + \gamma_1 \log_3 \frac{1}{\alpha_1} + \gamma_2 \log_3 \frac{1}{\alpha_2}.$$

It is clear that $\alpha$ determines a hyperplane that separate the space of all probability measures. Since we only consider the cases where $\gamma_0 + \gamma_1 + \gamma_2 = 1$, the above inequality simplifies to

$$\mathcal{H}(\alpha) \geq \gamma_0 \log_3 \frac{1}{\alpha_0} + \gamma_1 \log_3 \frac{1}{\alpha_1} + (1 - \gamma_0 - \gamma_1) \log_3 \frac{1}{1 - \alpha_0 - \alpha_1}.$$

Let $\gamma_0 = 0$, we may solve the above inequality and obtain the boundary point for $\alpha$ at $\gamma_0 = 0$ is $\gamma_1 = \frac{9}{25}$. Similarly, the boundary point for $\beta$ at $\gamma_0 = 0$ is approximately $\gamma_1 = 0.3965181711$.

Let $\gamma^* = (0, 0.37, 0.63)$.

$$\mathcal{H}(\alpha) - \mathrm{E}_{\gamma^*} \log_3 \frac{1}{\alpha} \approx -0.01154648767$$

and

$$\mathcal{H}(\beta) - \mathrm{E}_{\gamma^*} \log_3 \frac{1}{\beta} \approx 0.02593650702.$$

Thus $\beta$ dominates $\gamma^*$ but $\alpha$ does not dominate $\gamma^*$. □

**Lemma C.1.** *([14]). For every $n \geq m \geq 2$ and every partition $\vec{a} = (a_0, \ldots, a_{m-1})$ of $n$, there are more than*

$$m^{n\mathcal{H}_m(\frac{\vec{a}}{n}) - (m+1)\log_m n}$$

*strings $u$ of length $n$ and $\#(i, u) = a_i$ for each $i \in \Sigma_m$.*

**Theorem C.2.** *([8]). Let $d$ be an $s$-supergale, where $s \in [0, \infty)$. Then for all $w \in \Sigma_m^*$, $l \in \mathbb{N}$, and $0 < \alpha \in \mathbb{R}$, there are fewer than $\frac{m^l}{\alpha}$ strings $u \in \Sigma_m^l$ for which $d(wu) > \alpha m^{(s-1)l} d(w)$.*

**Proof of Theorem 5.7.** First we prove $\dim_\mathrm{H}(X) \geq H$. It suffices to show that for all $s < H$, $\dim_\mathrm{H}(X) \geq s$.

Let $s < H$. Let $d$ be an arbitrary $s$-supergale. Let $s' = (H + s)/2$. Let $n_0 \in \mathbb{N}$ be such that $\sqrt{m} < n_0(H - s')$ and $m^{s'n_0 - (m+1)\log_m n_0} > 2^{sn_0 + 1}$.

Fix an $S \in X$ such that $\liminf_{n \to \infty} \mathcal{H}_m(\vec{\pi}(S, n)) > s'$.

For each $i \geq n_0$, let $\{\vec{\beta}_{i,1}, \ldots, \vec{\beta}_{i,c_i}\} \subseteq \Delta(\Sigma_m)$ be such that for each $j \in [1..c_i]$, $\vec{\beta}_{i,j} = \frac{k}{n}$ for some $k \leq n$ and $\mathcal{H}_m(\vec{\beta}_{i,j}) > s'$; for all $\vec{\beta} \in F(S)$ there exists $j \in [1..c_i]$ such that $|\vec{\beta}_{i,j} - \vec{\beta}| < 1/i$; for all $j \in [1..c_i]$, there exists $\vec{\beta} \in F(S)$ such that $|\vec{\beta}_{i,j} - \vec{\beta}| < 1/i$; for all $j \in [1..c_i - 1]$, $|\vec{\beta}_{i,j} - \vec{\beta}_{i,j+1}| < \frac{1}{i}$; for all $i \geq n_0$, $|\vec{\beta}_{i+1,0} - \vec{\beta}_{i,c_i}| < \frac{1}{i+1}$.

Now, we first construct a sequence $S' \in \Sigma_m^\infty$ by building its prefixes inductively.

Let $w_0$ be such that $|w_0| = 2^{n_0}$. Note that the choice of $w_0$ does not affect the argument, since $w_0$ does not change the asymptotic behavior of the sequence. Without loss of generality, assume $\vec{\pi}(w_0, |w_0|) = \vec{\beta}_{n_0,1}$.

For all $n > 0$, assume $w_{n-1}$ is already constructed. Let $w_{n,0} = w_{n-1}$. We construct inductively $w_{n,1}, \ldots, w_{n,c_n}$ and then let $w_n = w_{n,c_n}$.

For $j > 0$, assume $w_{n,j-1}$ is already constructed.

Let $l = n_0 + n - 1$.

For each $l, j$, let
$$B_{l,j} = \left\{ u \in \Sigma_m^l \mid \vec{\pi}(u, l) = \vec{\beta}_{l,j} \right\}.$$

For each $l \geq n_0$ and $w \in \Sigma_m^*$, let
$$W_{l,w} = \left\{ u \in \Sigma_m^l \mid d(wu) \leq \frac{1}{m} d(w) \right\}.$$

Since $d$ is an $s$-supergale, by Theorem C.2, for all $w \in \Sigma_m^*$, there are fewer than $m^{sl+1}$ strings $u \in \Sigma_m^l$ for which $d(wu) > \frac{1}{m} d(w)$. By the choice of $n_0$, $\vec{\beta}_{l,j}$, and Lemma C.1,
$$|B_{l,j}| > m^{sl+1},$$
i.e., $W_{l,w} \cap B_{l,j} \neq \varnothing$.

Let $u_1 \in W_{l,w} \cap B_{l,j}$. For all $i \in [2..2^{|w_{n,j-1}|}]$, let $u_i \in W_{l,wu_1\ldots u_{i-1}} \cap B_{l,j}$.

Let $w_{n,j} = w_{n,j-1} u_1 \ldots u_{2^{|w_{n,j-1}|}}$.

Let $S' = \lim_{n \to \infty} w_n$.

Note that when $w_n$ is being constructed, $l \leq \lfloor \log_m |w_{n,j-1}| \rfloor$. It is then easy to verify that $S' \notin S^\infty[d]$.

Now we verify that $\Pi(S) = \Pi(S')$. Then $S' \in X$, since $X$ is defined by asymptotic frequency.

Let $\vec{\beta} \in \Pi(S)$ be arbitrary. For each $l = n_0 + n - 1$, there exists some $j_l$ such that $|\vec{\beta} - \vec{\beta}_{l,j_l}| < \frac{1}{l}$. Then by the construction,
$$|\vec{\pi}(w_{l,j_l}, |w_{l,j_l}|) - \vec{\beta}_{l,j_l}| < \sqrt{m} \frac{2}{|w_{l,j_l}|} < \frac{1}{l}.$$

So it is clear that
$$|\vec{\pi}(w_{l,j_l}, |w_{l,j_l}|) - \vec{\beta}| < \frac{2\sqrt{m}}{l}.$$

Thus
$$\lim_{l \to \infty} \vec{\pi}(w_{l,j_l}, |w_{l,j_l}|) = \vec{\beta}.$$

Since $w_{l,j_l} \sqsubseteq S'$ for all $l = n_0 + n - 1$. So we have for all $n \in \mathbb{N}$, $\vec{\beta} \in \bar{\Pi}_n(S')$, hence $\vec{\beta} \in \Pi(S')$. Therefore $\Pi(S) \subseteq \Pi(S')$.

Now, let $\vec{\beta} \notin \Pi(S)$. Since $\Pi(S)$ is closed, there exists $\delta > 0$ such that for all $\vec{\beta}' \in \Pi(S)$, $|\vec{\beta} - \vec{\beta}'| > \delta$.

Let $n_1$ be such that $l_1 = n_0 + n_1 - 1 > \frac{8m}{\delta}$. By construction, for all $l \geq l_1$, all $j \in [1..c_l]$, and all $|w_{l,j-1}| \leq k \leq |w_{l,j}|$,
$$|\vec{\pi}(w_{l,j}, |w_{l,j}|) - \vec{\pi}(w_{l,j}, k)| < \frac{2\sqrt{m}}{l}.$$

Also, for all $l \geq l_1$ and all $j \in [1..c_l]$, there exists $\vec{\beta}' \in \Pi(S)$ such that
$$|\vec{\pi}(w_{l,j}, |w_{l,j}|) - \vec{\beta}'| < \frac{2\sqrt{m}}{l}.$$

Thus for all $k > |w_{l_1,1}|$, there exists $\vec{\beta}' \in \Pi(S)$ such that
$$|\vec{\pi}(S, k) - \vec{\beta}'| < \frac{4m}{l}.$$

Therefore, for all $k > |w_{l_1,1}|$

$$|\vec{\pi}(S,k) - \vec{\beta}'| < \frac{4m}{l_1} < \frac{\delta}{2}.$$

Thus for all sufficiently large $k$,

$$|\vec{\pi}(S,k) - \vec{\beta}| > \frac{\delta}{2}.$$

So there exists $n_2 \in \mathbb{N}$ such that for all $n \geq n_2$, $\vec{\beta} \notin \bar{\Pi}_n$, i.e., $\vec{\beta} \notin \Pi(S')$.

Now we have that $S' \in X$. Since $S' \notin S^\infty[d]$, $s < H$ is arbitrary, and $d$ is an arbitrary $s$-supergale,

$$\dim_\mathrm{H}(X) \geq H.$$

By a similar construction, we may prove that

$$\dim_\mathrm{P}(X) \geq P.$$

In the following, we prove the finite-state dimension upper bounds. Given $\vec{\alpha} \in \Delta(\Sigma_m)$, define $B(\vec{\alpha}, r)$ as

$$B(\vec{\alpha}, r) = \Delta(\Sigma_m) \cap \left\{ \vec{\beta} \in \mathbb{R}^m \mid (\forall i)[\beta_i < \alpha_i m^r \text{ and } \beta_i > \alpha_i m^{-r}] \right\}.$$

Let

$$F(X) = \{\vec{\alpha} \in \Delta(\Sigma_m) \mid \mathcal{H}(\vec{\alpha}) = H\}.$$

Let $\epsilon > 0$. Let

$$\mathcal{C} = \left\{ B(\vec{\alpha}, \tfrac{\epsilon}{2}) \mid \vec{\alpha} \in F(X) \right\}.$$

It is clear that $\mathcal{C}$ is an open cover of $F(X)$. Since $F(X)$ is compact, there exists $C \subseteq \Delta(\Sigma_m)$ such that $|C| < \infty$ and

$$F(X) \subseteq \bigcup_{\vec{\alpha} \in C} B(\vec{\alpha}, \tfrac{\epsilon}{2}).$$

Let $S \in X$. Then $\liminf_{n \to \infty} \mathcal{H}_m(\vec{\pi}(S,n)) \leq H$. By Theorem 5.4, there exists $\vec{\alpha}^* \in F(X)$ such that $\vec{\alpha}^* \gg^\epsilon \vec{\pi}(S,n)$ for infinitely many $n \in \mathbb{N}$. Then by the construction of $C$, there exists $\vec{\alpha} \in C$ such that $\vec{\alpha}^* \in B(\vec{\alpha}, \tfrac{\epsilon}{2})$. Now, we have that for infinitely many $n \in \mathbb{N}$,

$$\mathcal{H}_m(\vec{\alpha}) = \mathcal{H}_m(\vec{\alpha}^*) \geq \mathrm{E}_{\vec{\pi}(S,n)} \log_m \frac{1}{\vec{\alpha}^*} - \frac{\epsilon}{2}$$

$$= \mathrm{E}_{\vec{\pi}(S,n)} \log_m \frac{1}{\vec{\alpha}} + \mathrm{E}_{\vec{\pi}(S,n)} \log_m \frac{\vec{\alpha}}{\vec{\alpha}^*} - \frac{\epsilon}{2}.$$

By the definition of $B(\alpha, \tfrac{\epsilon}{2})$,

$$\mathcal{H}_m(\vec{\alpha}) \geq \mathrm{E}_{\vec{\pi}(S,n)} \log_m \frac{1}{\vec{\alpha}} - \epsilon,$$

i.e., $\vec{\alpha} \gg^\epsilon \vec{\pi}(S,n)$ for infinitely many $n \in \mathbb{N}$. Since $S \in X$ is arbitrary, we may partition $X$ as $X = \bigcup_{\vec{\alpha} \in C} X_{\vec{\alpha}}$ such that for every $\vec{\alpha} \in C$,

$$X_{\vec{\alpha}} = \{S \in X \mid \vec{\alpha} \gg^\epsilon \vec{\pi}(S,n) \text{ for infinitely many } n \in \mathbb{N}\}.$$

Since $\epsilon > 0$ is arbitrary, thus by Theorem 5.5, $\dim_{\mathrm{FS}}(X_{\vec{\alpha}}) \leq \mathcal{H}_m(\vec{\alpha}) = H$ for every $\vec{\alpha} \in C$. Since $|C| < \infty$, by Theorem 3.2, $\dim_{\mathrm{FS}}(X) \leq H$. Similarly, $\mathrm{Dim}_{\mathrm{FS}}(X) \leq P$. □